%% file: main.tex
\documentclass{new_tlp}
\usepackage{mathptmx}

\providecommand\AMSLaTeX{AMS\,\LaTeX}
\newcommand\eg{\emph{e.g.}\ }
\newcommand\etc{\emph{etc.}}
\newcommand\bcmdtab{\noindent\bgroup\tabcolsep=0pt%
  \begin{tabular}{@{}p{10pc}@{}p{20pc}@{}}}
\newcommand\ecmdtab{\end{tabular}\egroup}
\newcommand\rch[1]{$\longrightarrow\rlap{$#1$}$\hspace{1em}}
\newcommand\lra{\ensuremath{\quad\longrightarrow\quad}}

  \title[A Stale Synchronous Model for Recursive Computation]
        {A Case for Stale Synchronous Distributed Model 
        for Declarative Recursive Computation}

  \author[A. Das and C. Zaniolo]
         {ARIYAM DAS and CARLO ZANIOLO\\
         Department of Computer Science, University of California, Los Angeles, USA\\
         \email{\{ariyam, zaniolo\}@cs.ucla.edu}}

\jdate{March 2003}
\pubyear{2003}
\pagerange{\pageref{firstpage}--\pageref{lastpage}}
\doi{S1471068401001193}

\let\proof\relax

\usepackage{mathptmx}
\usepackage{amsmath}
\usepackage{color}
\usepackage{amssymb, amsthm}
\usepackage{graphicx}
\usepackage[T1]{fontenc}
\usepackage{xspace}
\usepackage{cite}
\usepackage{amsthm}
\usepackage{enumerate}
\usepackage{algorithmic}
\usepackage[framemethod=TikZ]{mdframed}
\usepackage{multirow}
\usepackage{hyperref}

\def\prem{$\cal P$$\!reM$\xspace}

\newcommand{\bldl}{\smallskip\[\begin{array}{ll}}
\newcommand{\cldl}{\[\begin{array}{ll}}
\newcommand{\eldl}{\end{array}\]\rm}

\def\magg#1{min \langle #1 \rangle}

\newcommand{\arrow}{~\mbox{\tt <-}~}

\newcommand*{\revision}{\textcolor{black}} 

\newtheorem{definition}{Definition}
\newtheorem{example}{Example}
\newtheorem{theorem}{Theorem}

\newtheorem{innercustomgeneric}{\customgenericname}
\providecommand{\customgenericname}{}
\newcommand{\newcustomtheorem}[2]{%
  \newenvironment{#1}[1]
  {%
   \renewcommand\customgenericname{#2}%
   \renewcommand\theinnercustomgeneric{##1}%
   \innercustomgeneric
  }
  {\endinnercustomgeneric}
}

\newcustomtheorem{customthm}{Theorem}
\newcustomtheorem{customlemma}{Lemma}

\begin{document}
\nocite{*}

\label{firstpage}

\maketitle
\input{abstract.tex}

  \begin{keywords}
    Datalog, Deductive Databases, Recursive Query, Stale Synchronous Parallel Model, Bulk Synchronous Parallel Model, Parallel and Distributed Computing
  \end{keywords}


\input{introduction.tex}
\input{prem.tex}
\input{parallel.tex}
\input{premparallel.tex}
\input{relaxedmodels.tex}
\input{ssp.tex}
\input{conclusion.tex}


\bibliographystyle{acmtrans}
\bibliography{references}

\input{app.tex}

\label{lastpage}
\end{document}

%% file: abstract.tex
\begin{abstract} 
%
A large class of traditional graph and data mining algorithms can be concisely expressed in Datalog, and other Logic-based languages, once aggregates  are allowed in recursion.  In fact, for most BigData algorithms, the difficult semantic issues raised by the use of  non-monotonic aggregates in recursion are solved by \emph{Pre-Mappability} (\prem), a property that assures that for a program with aggregates in recursion there is an equivalent aggregate-stratified program. In this paper we show that, by bringing together  the formal abstract  semantics of stratified programs with the efficient operational one of unstratified programs, \prem  can also facilitate and improve their parallel execution. We  prove that \prem-optimized lock-free \revision{and decomposable} parallel semi-naive evaluations  produce the same results as the single executor programs.
Therefore, \prem can be assimilated into the data-parallel computation plans of different distributed systems, irrespective of whether these follow bulk synchronous parallel (BSP) or asynchronous computing models.  
In addition, we show that non-linear recursive queries can be evaluated using a hybrid stale synchronous parallel (SSP) model on distributed environments. After providing a formal correctness proof for the recursive query evaluation with \prem under this relaxed synchronization model, we present experimental evidence of its benefits.
%
This paper is under consideration for acceptance in Theory and Practice of Logic Programming (TPLP).
\end{abstract}

%% file: introduction.tex
\section{Introduction}\label{intro}
The  growing interest in Datalog-based declarative systems like \emph{LogicBlox} \citeS{logicblox}, \emph{BigDatalog} \citeS{bigdatalog}, \emph{SociaLite} \citeS{socialite}, \emph{BigDatalog-MC} \citeS{bigdatalog-mc} and \emph{Myria} \citeS{Myria} has brought together important advances on two fronts: 
(i) Firstly,  Datalog, with support for aggregates in recursion  \citeS{monotonic_agg}, has sufficient  power to express succinctly  declarative applications ranging from complex graph queries to advanced data mining tasks,  such as frequent pattern mining and decision tree induction \citeS{scalingup-tplp2018}. 
(ii) Secondly,  modern  architectures supporting in-memory parallel and distributed computing can deliver scalability and performance for this new generation of Datalog systems. 

For example \emph{BigDatalog} (\revision{bulk synchronous parallel} processing on shared-nothing architecture), 
\emph{BigDatalog-MC} (lock-free parallel processing on shared-memory multicore architecture), 
\emph{Myria} (asynchronous processing on shared-nothing architecture) spearheaded the system-level scheduling, planning and optimization for different parallel computing models. This line of work was quite successful for Datalog, and also for recursive SQL queries that have borrowed this technology \citeS{rasql}). Indeed, our recent general-purpose Datalog systems surpassed commercial graph systems like GraphX on many classical graph queries  in terms of performance and scalability \citeS{bigdatalog}. 

\revision{Much of the theoretical groundwork contributing to the success of these parallel Datalog systems was laid out in the 90s. For example,} 
in their foundation work  \citeS{GangulyParallel}  investigated parallel \emph{coordination-free} (asynchronous) bottom-up evaluations of simple linear recursive programs (without any aggregates). 
In fact, many recent works have pushed this idea forward under the broader umbrella of 
\emph{CALM conjecture} (Consistency And Logical Monotonicity) \citeS{CalmConjecture} which establishes that monotonic Datalog (Datalog without negation or aggregates) programs 
can be computed in an \emph{eventually consistent, coordination-free} manner \citeS{AmelootCalm2}, \citeS{AmelootCalmRefined}. This line of work led to the asynchronous data-parallel (for \emph{Myria}) and lock-free evaluation plans for many of the aforementioned systems (e.g. \emph{BigDatalog-MC}).    
Simultaneously, another branch of research about `parallel correctness' for simple non-recursive conjunctive queries \citeS{ParallelCorrectness} focused on optimal data distribution policies for re-partitioning the initial data under \revision{Massively Parallel Communication} model (MPC). 
However, notably,  this theoretical groundwork left out programs using aggregates in recursion, for which the existence of a formal semantics could not be guaranteed. But, this situation has changed recently because of the introduction of the notion  of \emph{Pre-Mappability}\footnote{\revision{In our initial work \citeS{zaniolo-tplp2017}, we interchangeably used the term \emph{Pre-Applicability}. However, in our follow-up works \citeS{scalingup-tplp2018}, \citeS{zaniolo-amw2018}, we consistently used the term \emph{Pre-Mappability} since the latter was deemed more appropriate in the context of `pre-mapping' aggregates and constraints to recursive rules.}} (\prem) \citeS{zaniolo-tplp2017} that has made possible the use of  aggregates in recursion to express efficiently a large range of applications \citeS{scalingup-tplp2018}.  
\revision{A key aspect of this line of work has been the use of non-monotonic aggregates and pre-mappable constraints inside recursion, while preserving the  formal declarative semantics of aggregate-stratified
programs, thanks to the notion of \prem that guarantees their equivalence.
Unlike more complex non-monotonic semantics, stratification 
is a syntactic condition that is easily checked by users (and compilers), who 
know that the presence of a formal declarative semantics guarantees
the portability of their applications over multiple platforms.
Furthermore, evidence is mounting that a higher potential for
parallelism is also gained under \prem. 
Naturally, we would like to examine the applicability of \prem under a parallel and distributed setting and analyze 
its potential gains using the rich models of parallelism previously proposed for Datalog and other logic systems.
}

%
%
%
%
In this paper, therefore, we begin by examining how \prem interacts under \revision{a} parallel setting, and address the question of whether it can be incorporated into the parallel evaluation plans on shared-memory and shared-nothing architectures. Furthermore, the current crop of Datalog systems supporting aggregates in recursion have only explored \revision{Bulk Synchronous Parallel (BSP)} and asynchronous distributed computing models. However, the new emerging paradigm of Stale Synchronous Parallel (SSP) processing model \citeS{SSPCui} has shown to speed up big data analytics and machine learning algorithm execution on distributed environments \citeS{SSPPetuum}, \citeS{SSPMuli} with \emph{bounded staleness}. 
\revision{SSP processing allows each worker in a distributed setting to see and use another worker's obsolete (stale) intermediate solution, which is out-of-date only by a limited (bounded) number of epochs. On the contrary, in a BSP model every worker coordinates at the end of each round of computation and sees each others' current intermediate results. This relaxation of the synchronization barrier in a SSP model can reduce idle waiting of the workers (time spent waiting to synchronize), particularly when one or more workers (stragglers) lag behind others in terms of computation. 
Thus, in this paper, we also explore if declarative recursive computation can be executed under the loose consistency model of SSP processing and if it has the same convergence as that under a BSP processing framework.}
To our surprise, we find \prem dovetails excellently with SSP model for a class of non-linear recursive queries with aggregates, which  are not embarrassingly parallel and still require some coordination between the workers to reach eventual consistency \citeS{interlandi_tanca_2018}. Thus\revision{,} the contributions of this paper can be summarized as follows: 

\begin{itemize}
    \item  We show that \prem is applicable to parallel bottom-up semi-naive evaluation plan, terminating at the same minimal fixpoint as the corresponding single executor based sequential execution.
    \item We further show how recursive query evaluation with \prem can operate effectively under a  SSP distributed model. 
    \item Finally, we discuss the merits and demerits of a SSP model with initial empirical results on some recursive query examples, thus opening up an interesting direction for future research.
\end{itemize}

%
%
%
%

%% file: prem.tex
 \section{An Overview of \prem}\label{prem-overview}
This section provides a brief overview about \prem and some of its properties \citeS{zaniolo-amw2016}, \citeS{zaniolo-amw2018}.
Consider the Datalog query in Example \ref{allpairs} that computes the 
shortest path between all pairs of vertices in a graph, given by the relation 
 {\tt arc(X, Y, D)}, where {\tt D} is the distance between source node {\tt X} and destination node {\tt Y}. The {\tt min$\langle$D$\rangle$} syntax in our example indicates $min$ aggregate on the cost variable {\tt D}, while {\tt (X, Y)} refer to the group-by arguments. This head notation for aggregates directly follows from SQL-2 syntax, where cost argument for the aggregate consists of one variable and group-by arguments can have zero or more variables.  
 Rules \revision{$r_{1.3}$} in the example shows that the aggregate 
 $min$ is computed at a stratum higher than the recursive rule (\revision{$r_{1.2}$}).
 \begin{example}{All Pairs Shortest Path}
\label{allpairs}
\begin{displaymath}
\begin{aligned}
& r_{1.1}: {\tt path(X, Y, D)} \arrow {\tt arc(X, Y, D)}. \\
& r_{1.2}: {\tt path(X, Y, D)} \arrow {\tt path(X, Z, Dxz)}, {\tt arc(Z, Y, Dzy)},  {\tt D=Dxz+Dzy}. \\
& r_{1.3}: {\tt shortestpath(X, Y, \magg{D})} \arrow {\tt path(X, Y, D)}. \\
\end{aligned}
\end{displaymath}
\end{example}

Incidentally, \revision{$r_{1.3}$} can also be expressed with stratified negation as shown in rules \revision{$r_{1.4}$} and \revision{$r_{1.5}$}. This guarantees that the program has a perfect-model semantics, although an iterated fixpoint computation of it can be very inefficient and even non-terminating in presence of cycles.
\begin{displaymath}
\begin{aligned}
& r_{1.4}: {\tt shortestpath(X, Y, D)} \arrow {\tt path(X, Y, D)}, {\tt \neg betterpath(X, Y, D)}. \\
& r_{1.5}: {\tt betterpath(X, Y, D)} \arrow \revision{{\tt path(X, Y, D)},} {\tt path(X, Y, Dxy)}, {\tt Dxy < D}. \\
\end{aligned}
\end{displaymath}

\noindent \textbf{\prem Application.} The aforementioned inefficiency can be mitigated with \prem, if the $min$ aggregate can be pushed inside the fixpoint computation, as shown in rules \revision{$r_{2.1}$} and \revision{$r_{2.2}$}. The following program under \prem has a stable model semantics and \citeS{scalingup-tplp2018} showed that this transformation is indeed equivalence-preserving with an assured convergence to a minimal fixpoint within a finite number of iterations.
In other words, without \prem the shortest path in our example (according to rule \revision{$r_{1.3}$}) is given by the subset of the minimal model (computed from rules \revision{$r_{1.1}, r_{1.2}$}) obtained after removing {\tt path} atoms that did not satisfy the $min$ cost constraint for a given source-destination pair. However, with \prem, the transfer of $min$ cost constraint inside recursion results in an optimized program, where \revision{the} fixpoint computation is performed more efficiently, eventually achieving the \revision{same shortest path values (as those produced in the perfect model of the earlier program)} by simply copying the atoms from {\tt path} under the name {\tt shortestpath} \revision{(rule $r_{2.3}$)} after the least fixpoint computation terminates.
\begin{displaymath}
\begin{aligned}
& r_{2.1}: {\tt path(X, Y, \magg{D})} \arrow {\tt arc(X, Y, D)}. \\
& r_{2.2}: {\tt path(X, Y, \magg{D})} \arrow {\tt path(X, Z, Dxz)}, {\tt arc(Z, Y, Dzy)},  {\tt D=Dxz+Dzy}. \\
& r_{2.3}: {\tt shortestpath(X, Y, D)} \arrow {\tt path(X, Y, D)}. \\
\end{aligned}
\end{displaymath}

\noindent \textbf{Formal Definition of \prem.} For a given Datalog program, let $P$ be the rules defining a (set of mutually) recursive predicate(s) and $T$ be the corresponding \emph{Immediate Consequence Operator} (ICO) defined over $P$. Then, a constraint $\gamma$ is said to be \prem to $T$  (and to $P$) when, for every interpretation $I$ of the program, we have $\gamma(T(I)) = \gamma (T(\gamma (I)))$. 

\revision{In Example \ref{allpairs}, the final rule $r_{1.3}$ imposes the constraint $\tt{\gamma = (X, Y, \magg{D})}$ on $I = \tt{path(X, Y, D)}$ (representing all possible paths) to eventually yield the shortest path between all pairs of nodes. Thus, the aggregate-stratified program defined by rules $r_{1.1}-r_{1.3}$ is equivalent to $\gamma(T(I))$ in the definition of \prem.
On the other hand, with $min$ aggregate pushed inside recursion, recursive rules $r_{2.1}-r_{2.2}$ represent $\gamma(T(\gamma (I)))$.} \\

\noindent \textbf{\prem Properties.} We now discuss some important results about \prem from \citeS{zaniolo-tplp2017}. We refer interested readers to our paper \citeS{zaniolo-tplp2017} for the detailed proofs.
Let $T_\gamma$ denote the \emph{constrained immediate consequence operator}, where constraint $\gamma$ is applied after the ICO $T$, i.e., $T_\gamma(I) = \gamma(T(I))$. 
The following results hold when $\gamma$ is \prem to a positive program $P$ with ICO $T$:
\begin{enumerate}[(i).]
    \item If $I = T(I)$ is a fixpoint for $T$, then $I' = \gamma(I)$ is a fixpoint for $T_\gamma(I)$, i.e., $I' = T_\gamma(I')$.
    \item For some integer $n$, if $T_\gamma^{\uparrow n}(\emptyset) = T_\gamma^{\uparrow n+1}(\emptyset)$, then $T_\gamma^{\uparrow n}(\emptyset) = T_\gamma^{\uparrow n+1}(\emptyset)$ is a minimal fixpoint for $T_\gamma$ and 
    $T_\gamma^{\uparrow n}(\emptyset) = \gamma(T^{\uparrow \omega}(\emptyset))$, 
    where $T^{\uparrow \omega} = \bigcup\limits_{n \geq 1} T^{\uparrow n}$
\end{enumerate}

\noindent \textbf{\prem Provability.}
We can verify if \prem holds for a recursive rule by explicitly validating 
$\gamma(T(I)) = \gamma (T(\gamma (I)))$, i.e., $T_\gamma(I) = T_\gamma(\gamma(I))$ 
at every iteration of the fixpoint computation. 
To simplify, this would indicate that we can verify if the $min$ constraint can be pushed inside recursion in rule \revision{$r_{2.2}$} by inserting an additional goal $is\_min$ in the body of the rule as follows: 
\begin{displaymath}
\begin{aligned}
& r'_{2.2}: {\tt path(X, Y, \magg{D})} \arrow {\tt path(X, Z, Dxz)}, \footnotesize{{\tt {\bf is\_min}((X, Z), Dxz)}}, {\tt arc(Z, Y, Dzy)},  {\tt D=Dxz+Dzy}. \\
\end{aligned}
\end{displaymath}
This additional goal in the body pre-applies the constraint $\gamma$ on $I$, followed by the application of $T_\gamma$ operator, i.e., it expresses $T_\gamma(\gamma(I))$.  
Note, the $is\_min$ constraint is satisfied by {\tt Dxz}, if it is the minimum value seen yet in the fixpoint computation for the source-destination pair {\tt (X, Z)}. It is also evident that any other distance value between {\tt (X, Z)}, which violates the $is\_min$ constraint, will also not satisfy the $min$ aggregate at the head of the rule, since the additional goal minimizes the sum {\tt D} for each {\tt Dzy}. Thus, this new goal in the body does not alter the ICO mapping defined by the original recursive rule, thereby proving $\gamma$ is \prem in this example program. More broadly speaking, these additional goals can be formally defined as ``half functional dependencies'', borrowing the terminology from classical database theory of Functional and Multi-Valued Dependencies (FDs and MVDs). 
We next present the formal definition of \emph{half FD} from \citeS{zaniolo-amw2018}, which will be used later for our proofs. 
\begin{definition}{(Half Functional Dependency).}
Let $R(\Omega)$ be a relation on a set of attributes $\Omega$, $X \subset \Omega$ 
and $A \in \Omega - X$.
Considering the domain of $A$ to be totally ordered,
a tuple $t \in R$ is said to satisfy the $min$-constraint $is\_min((X),A)$ (denoted as  $X\stackrel{min}{\rightharpoonup}A$),  when  $R$  contains  no  tuple  with  the  same  $X$-value  and a smaller $A$-value. Similarly, a tuple $t \in R$ satisfies a $max$-constraint $is\_max((X),A)$ (denoted as $X\stackrel{max}{\rightharpoondown}A$)  if $R$ has no tuple with the same $X$-value and a $larger$ $A$-value.
\end{definition}
For any $min$ or $max$ constraint to be \prem to a positive program $P$, the corresponding half FD should hold for the relational view of the relevant recursive predicate across every interpretation $I$ of $P$, where a relational view for predicate $q$ is defined as $R_q=\{(x_1,...,x_n)|q(x_1,...,x_n) \in I\}$ for a given $I$.
\citeS{zaniolo-amw2018} provides generic templates, based on Functional and Multi-valued Dependencies, for identifying constraints that satisfy \prem.\\

\noindent \textbf{\prem with Semi-Naive Evaluation.}
A naive fixpoint computation trivially generates new atoms from the entire set of atoms available at the end of the last fixpoint iteration. Semi-naive evaluation improves over this naive fixpoint computation with the aid of the following enhancements:
\begin{enumerate}
    \item At every iteration, track \emph{only} the new atoms produced. 
    \item Rules are re-written into their differential versions, so that only new atoms are produced and old atoms are never generated redundantly.
    \item Ensure step (2) does not generate any duplicate atoms.
\end{enumerate}
For programs where \prem can be applied, steps (1) and (2) remain identical. However, step (3) is extended so that (i) new atoms produced may not be retained, if they do not satisfy the constraint $\gamma$ and (ii) existing atoms may get updated and thereafter tracked for the next iteration. For example, new atoms produced from rule $r_{2.2}$ are added to the working set and tracked only if a new source-destination ({\tt X,Y}) path is discovered. On the other hand, if the new {\tt path} atom, thus produced, has a smaller distance than the one in the working set, then the distance of the existing {\tt path} atom is updated to satisfy the $min$-constraint. However, if new {\tt path} atoms are generated, which have larger distances, then they are simply ignored. This understanding of \prem for semi-naive evaluation leads to a case for SSP model, where significant communication can be saved by condensing multiple updates into one. This is discussed in detail later in Section \ref{relaxedsync}.

%
%

%% file: parallel.tex
\section{An Overview of Parallel Bottom-Up Evaluation}\label{parallel-overview}
One of the early foundational works that established a standard technique to parallelize bottom-up evaluation of \emph{linear recursive} queries was presented in \citeS{GangulyParallel}. 
The authors proposed a \emph{substitution partitioned parallelization} scheme, where  
\revision{the} set of possible ground substitutions\revision{, i.e.,} the base (extensional database) and derived relation (intensional database) atoms in the Datalog program are disjointedly partitioned, using a hash-based discriminating function, so that each partition of possible ground substitutions is mapped to exactly one of the parallel workers. 
The entire computation is then divided among all the workers, operating in parallel, where each worker only processes the partition of ground substitutions mapped to it during the bottom-up semi-naive evaluation. Since, each worker operates on a distinct non-overlapping partition of ground substitutions, no two workers perform the same or redundant computation\revision{, i.e.,} this scheme is \emph{non-redundant}. 
Formally, if $v(r)$ is a non-repetitive sequence of variables appearing in the body of rule $r$ and $\mathcal{W}$ denotes a finite set of parallel workers, then  $h: v(r) \longrightarrow \mathcal{W}$ is a discriminating hash function that divides the workload by assigning the ground substitution and corresponding processing to exactly one worker. The workers can send and receive information (ground instances from partially computed derived relations) to and from other workers to finish the assigned computation tasks.
Ganguly et al. summarized the correctness of this parallelization scheme with the following result:

%

%
%
\vspace{6pt}

\noindent \textbf{Correctness of Partitioned Parallelization Scheme.} 
Let $P$ be a recursive Datalog program to be executed over $\mathcal{W}$ workers. Under the partitioned parallelization scheme, let $Q_i$ be the program to be executed at worker $i$ and let $Q = \bigcup\limits_{1 \leq i \leq \mathcal{W}} Q_i$. 
Then, for every interpretation, the least model of the recursive relation in $Q$ is identical to the least model obtained from the sequential execution of $P$. 

Note, the above parallelization strategy did not involve aggregates in recursion. But, nevertheless it was of significant consequence, since the scheme has been extended to derive \emph{lock-free parallel} plans for shared-memory architectures as well as 
\emph{sharded data parallel \revision{decomposable}} plans for shared-nothing distributed architectures to 
parallelize bottom-up semi-naive evaluation of Datalog programs. We discuss them next with examples. 

\vspace{6pt}

\noindent \textbf{Shared-Memory Architecture.}
A trivial \revision{hash-based} partitioning, as described above, can often lead to conflicts between different workers on a shared-memory architecture\footnote{For example, two distinct workers may update a {\tt path} atom for the same ({\tt X,Y}) pair in rule \revision{$r_{1.2}$}, if the hashing is done based on the ground instances of the sequence \{{\tt X,Z,Dxz,Z,Y,Dzy}\} or even on the sequence \{{\tt X,Z,Y}\}.}. This can be prevented with the implementation of classical locks to resolve read-write conflicts. However, recently, \citeS{yang-iclp2015} proposed a hash partitioning strategy based on \emph{discriminating sets} that allows \emph{lock-free} parallel evaluation of a broad class of generic queries including non-linear queries. We illustrate this with our running all pairs shortest path example. 

Assume the relations {\tt arc, path} and {\tt shortestpath} from example \ref{allpairs} (rules \revision{$r_{1.1}-r_{1.3}$}) are partitioned by the first column\footnote{The first attribute forms a \emph{discriminating set} that is used for partitioning.} (i.e.\revision{,} the source vertex), using a hash function $h$ that maps the source vertex to an integer between 1 to $\mathcal{W}$, latter denoting the number of workers. Now, a worker $i$ can execute the following program in parallel: 

\begin{displaymath}
\begin{aligned}
& r_{3.1}: {\tt path(X, Y, D)} \arrow {\tt arc(X, Y, D)}, {\tt h(X) = i}. \\
& r_{3.2}: {\tt path(X, Y, D)} \arrow {\tt path(X, Z, Dxz)}, {\tt arc(Z, Y, Dzy)},  {\tt D=Dxz+Dzy}, {\tt h(X) = i}. \\
& r_{3.3}: {\tt shortestpath(X, Y, \magg{D})} \arrow {\tt path(X, Y, D)}, {\tt h(X) = i}. \\
\end{aligned}
\end{displaymath}

\begin{enumerate}
    \item The $i^{th}$ worker executes rule \revision{$r_{3.1}$} by reading from the $i^{th}$ partition of arc.
    \item Once all the workers finish step (1), the $i^{th}$ worker begins semi-naive evaluation with rule \revision{$r_{3.2}$}, where it reads from the $i^{th}$ partition of {\tt path}, joins with the corresponding atoms from the {\tt arc} relation, which is shared across all the workers, and then writes new atoms into the same $i^{th}$ partition of {\tt path}.
    \item Once all the workers finish step (2), the semi-naive evaluation proceeds to the next iteration and repeats step (2) till the least fixpoint is reached. 
    \item In the final step, the $i^{th}$ worker computes the {\tt shortestpath} for the $i^{th}$ partition.
    \item All the {\tt shortestpath} data pooled across the workers produce the final query result. 
\end{enumerate}

It is easy to observe that the above parallel execution does not require any locks, since each worker is writing to exactly one partition and no two workers are writing to the same partition. We formally define the lock-free parallel bottom-up evaluation scheme next. 

\begin{definition}{(Lock-free Parallel Bottom-up Evaluation).}
Let $P$ be a recursive Datalog program to be executed over $\mathcal{W}$ workers and let $T$ be the corresponding ICO for the sequential execution of $P$. Under the lock-free parallel plan executed over $\mathcal{W}$ workers, let $Q_i$ be the program to be executed at worker $i$, producing an interpretation $I_i$ of the recursive predicate with the corresponding ICO $T_i$. 
Then, for every input of base relations, we have, 
$T_i^{\uparrow \omega}(\emptyset) \bigcap T_j^{\uparrow \omega}(\emptyset) = \emptyset$ for $1 \leq i,j \leq \mathcal{W}$, $i \ne j$. 
It also follows from the correctness of partitioned parallelization scheme that 
$\bigcup\limits_{1 \leq i \leq \mathcal{W}} T_i^{\uparrow \omega}(\emptyset) = T^{\uparrow \omega}(\emptyset)$. 
\end{definition}
The underlying strategy of a lock-free parallel plan to use disjointed data partitions have also been adopted to execute data-parallel distributed bottom-up evaluations, as explained next.

\vspace{6pt}

\noindent \textbf{Shared-Nothing Architecture.} Distributed systems like \emph{BigDatalog} \citeS{bigdatalog} also divide the entire dataset into disjointed data shards in an identical manner as the lock-free partitioning technique described above. Each data shard resides in the memory of a worker and this partitioning scheme reduces the data shuffling required across different workers \citeS{bigdatalog}. In the context of shared-nothing architecture, this sharding scheme and subsequent distributed bottom-up evaluation is termed as a \emph{decomposable plan} \citeS{bigdatalog}, \citeS{rasql}. In the rest of this paper, \revision{we will use the term `lock-free parallel plan' in the context of shared-memory architecture and `parallel decomposable plan' in the context of distributed environment for clarity.}

Distributed systems like \emph{BigDatalog} and \emph{SociaLite} \citeS{socialite} perform the fixpoint computation under BSP model with synchronized iterations. However, note that, if each node caches the {\tt arc} relation, then each node can operate independently without any \emph{co-ordination} or \emph{synchronization} with other nodes (i.e.\revision{,} step 3 listed before in the lock-free evaluation plan becomes unnecessary). The \emph{Myria} system follows this asynchronous computing model for the query evaluation. Interestingly, \citeS{GangulyParallel} showed that only a subclass of linear recursive queries\footnote{The dataflow graph corresponding to the linear recursive query must have a cycle.} can be executed in a \emph{co-ordination free} manner or asynchronously. Thus, for a large class of non-linear and even many linear recursive queries (e.g. same generation query \citeS{GangulyParallel}), BSP computing model has been the only viable option. 

%

%
%

%% file: premparallel.tex
\section{Parallel Evaluation with \prem}\label{premparallel}
  
In this section, we now examine if \prem can be easily integrated into \revision{the \emph{lock-free parallel} and \emph{parallel decomposable bottom-up evaluation} plans that have been} widely adopted across shared-memory and shared-nothing architectures for a broad range of generic queries. We next provide some interesting theoretical results. \\

\noindent \textbf{Lemma 1.} 
Let $R(\Omega)$ be a relation defined over a set of attributes $\Omega$, where $X \subset \Omega$ 
and $A \in \Omega - X$. For a subset $S$ of $X$ ($S \subseteq X $), if $R$ is divided into $k$ disjoint subsets $R_1, R_2, ..., R_k$ using a hash function $h: S \rightarrow k$ such that 
$R_i$ is defined as $R_i = \{e | e \in R \wedge h(e\left[ S \right]) = i\}$, then a tuple $t \in R$ satisfying $X\stackrel{min}{\rightharpoonup}A$ (or, $X\stackrel{max}{\rightharpoondown}A$) will also satisfy $X\stackrel{min}{\rightharpoonup}A$ (or, $X\stackrel{max}{\rightharpoondown}A$ respectively) 
over $R_i$ and vice versa, where $h(t\left[ S \right]) = i$.

\begin{proof*}
This follows directly from the fact that since $S \subseteq X$, for any two tuples $t_1, t_2 \in R$, if $t_1\left[ X \right] = t_2\left[ X \right]$, then 
$t_1\left[ S \right] = t_2\left[ S \right]$, i.e., any two tuples with the same $X$-value will be mapped into the same partition, decided by their common $S$-value. Since, all tuples with the same $X$-value belong to a single partition, any tuple $t \in R_i$ will satisfy 
$X\stackrel{min}{\rightharpoonup}A$ (or, $X\stackrel{max}{\rightharpoondown}A$) over both $R$ and $R_i$. 
\end{proof*}

\begin{theorem}\label{theorem-prem}
Let $P$ be a recursive Datalog program, $T$ be its corresponding ICO and let the constraint $\gamma$ be \prem to $T$ and $P$, resulting in the constrained ICO $T_{\gamma}$. Let $P$ be executed over $\mathcal{W}$ workers under a lock-free parallel \revision{(or parallel decomposable)} bottom-up evaluation plan, where $Q_i$ is the program executed at worker $i$ and $T_i$ be the corresponding ICO defined over $Q_i$. 
If the group-by arguments used for the $\gamma$ constraint also contain the discriminating set used for partitioning in the lock-free parallel \revision{(or parallel decomposable)} plan, then:
\end{theorem}

\begin{enumerate}[(i).] 
    \item \emph{$\gamma$ is also \prem to $T_i$ and $Q_i$, for $1 \leq i \leq \mathcal{W}$}.
    \item \emph{For some integer $n$, if  $T_\gamma^{\uparrow n}(\emptyset)$ is the minimal fixpoint for $T_{\gamma}$, then  
    $T_\gamma^{\uparrow n}(\emptyset) = \bigcup\limits_{1 \leq i \leq \mathcal{W}} T_{i_{\gamma} }^{\uparrow n}(\emptyset)$, where $T_{i_{\gamma}}$ denotes the constrained ICO with respect to $T_i$.}
\end{enumerate}

\begin{proof*}
The proof for (i) follows trivially from lemma 1 and the \prem provability technique discussed earlier in Section \ref{prem-overview}.

Since, $\gamma$ is \prem to $T$ and $P$,  $T_\gamma^{\uparrow n}(\emptyset) = \gamma(T^{\uparrow \omega}(\emptyset))$ according to the properties of \prem. 
Similarly, since for $1 \leq i \leq \mathcal{W}$, $\gamma_i$ is \prem to $T_i$ and $Q_i$ (from (i) of theorem \ref{theorem-prem}), $T_{i_{\gamma}}^{\uparrow n_i}(\emptyset) = \gamma(T_i^{\uparrow \omega}(\emptyset))$, for some integer $n_i$, where $T_{i_{\gamma}}^{\uparrow n_i}(\emptyset) = T_{i_{\gamma}}^{\uparrow (n_i+1)}(\emptyset)$ is the minimal fixpoint for $T_{i_{\gamma}}$.
Thus, $\bigcup\limits_{1 \leq i \leq \mathcal{W}} T_{i_{\gamma}}^{\uparrow n_i}(\emptyset)
= \bigcup\limits_{1 \leq i \leq \mathcal{W}} \gamma(T_i^{\uparrow \omega}(\emptyset))$. 
Now, $\gamma$ constraints are also trivially \prem to union over disjoint sets \citeS{zaniolo-tplp2017}, i.e., 
$\gamma(\bigcup\limits_{1 \leq i \leq \mathcal{W}} T_i^{\uparrow \omega}(\emptyset))
= \bigcup\limits_{1 \leq i \leq \mathcal{W}} \gamma(T_i^{\uparrow \omega}(\emptyset))$. 
Also recall from the definition of lock-free parallel \revision{(or parallel decomposable)} plan that 
$\bigcup\limits_{1 \leq i \leq \mathcal{W}} T_i^{\uparrow \omega}(\emptyset) = T^{\uparrow \omega}(\emptyset)$. 
Combining these aforementioned equalities, we get,\\
$T_\gamma^{\uparrow n}(\emptyset) = 
\gamma(T^{\uparrow \omega}(\emptyset)) =
\gamma(\bigcup\limits_{1 \leq i \leq \mathcal{W}} T_i^{\uparrow \omega}(\emptyset)) =
\bigcup\limits_{1 \leq i \leq \mathcal{W}} \gamma(T_i^{\uparrow \omega}(\emptyset)) =
\bigcup\limits_{1 \leq i \leq \mathcal{W}} T_{i_{\gamma}}^{\uparrow n_i}(\emptyset)$. \\
Since, $T_{i_{\gamma}}^{\uparrow n_i}(\emptyset)$ is the minimal fixpoint with respect to 
$T_{i_{\gamma}}$, for $n > n_i$, 
$T_{i_{\gamma}}^{\uparrow n}(\emptyset) = T_{i_{\gamma}}^{\uparrow n_i}(\emptyset)$. 
Therefore, $T_\gamma^{\uparrow n}(\emptyset) = \bigcup\limits_{1 \leq i \leq \mathcal{W}} T_{i_{\gamma} }^{\uparrow n}(\emptyset)$. 
\end{proof*}

Thus, following theorem \ref{theorem-prem}, we can push the $min$ constraint within the parallel recursive plan expressed by rules \revision{$r_{3.1}-r_{3.3}$} and rewrite them for worker $i$ as follows: 

\begin{displaymath}
\begin{aligned}
& r_{4.1}: {\tt path(X, Y, \magg{D})} \arrow {\tt arc(X, Y, D)}, {\tt h(X) = i}. \\
& r_{4.2}: {\tt path(X, Y, \magg{D})} \arrow {\tt path(X, Z, Dxz)}, {\tt arc(Z, Y, Dzy)},  {\tt D=Dxz+Dzy}, {\tt h(X) = i}. \\
& r_{4.3}: {\tt shortestpath(X, Y, D)} \arrow {\tt path(X, Y, D)}, {\tt h(X) = i}. \\
\end{aligned}
\end{displaymath}

Thus, we observe that pre-mappable constraints can be also easily pushed inside parallel lock-free \revision{(or parallel decomposable)} evaluation plans of recursive queries to yield the same minimal fixpoint, yet making them computationally more efficient and safe. Thus, \prem can be easily incorporated into the parallel computation plans (equivalent to rules \revision{$r_{4.1} - r_{4.3}$}) of different systems like \emph{BigDatalog-MC}, \emph{BigDatalog} and \emph{Myria}, irrespective of whether they use (1) shared-memory or shared-nothing architecture, or (2) they follow BSP or asynchronous computing models. 

%% file: relaxedmodels.tex
\section{A Case for Relaxed Synchronization}\label{relaxedsync}

We now consider a non-linear query, which is equivalent to the linear all pairs shortest path program with the application of \prem (rules \revision{$r_{2.1}-r_{2.3}$}). Since this is a non-linear query (rules \revision{$r_{5.1}-r_{5.3}$}), this program \emph{cannot} be executed in a \emph{coordination-free} manner or \emph{asynchronously} following the technique described in \citeS{GangulyParallel}. 

\begin{displaymath}
\begin{aligned}
& r_{5.1}: {\tt path(X, Y, \magg{D})} \arrow {\tt arc(X, Y, D)}. \\
& r_{5.2}: {\tt path(X, Y, \magg{D})} \arrow {\tt path(X, Z, Dxz)}, {\tt path(Z, Y, Dzy)},  {\tt D=Dxz+Dzy}. \\
& r_{5.3}: {\tt shortestpath(X, Y, D)} \arrow {\tt path(X, Y, D)}. \\
\end{aligned}
\end{displaymath}

\vspace{5pt}

However, as shown in \citeS{yang-iclp2015}, a simple query rewriting technique can produce an equivalent parallel \revision{decomposable} evaluation plan for this non-linear query. Rules \revision{$r_{6.1}-r_{6.4}$} show the equivalent \revision{decomposable} program, which can be executed by worker $i$ on a distributed system following a bulk synchronous parallel model. In this following \revision{decomposable} evaluation plan, there is a mandatory synchronization step (rule \revision{$r_{6.3}$}), where each worker $i$ (operating on the $i^{th}$ partition) copies the new atoms or updates in {\tt path} produced during the semi-naive evaluation from rule \revision{$r_{6.2}$} to  
{\tt path}$^{(1)}$ and the new {\tt path}$^{(1)}$ is then sent to other workers so that they can use it in the evaluation of rule \revision{$r_{6.2}$} in the next iteration.

\begin{displaymath}
\begin{aligned}
& r_{6.1}: {\tt path(X, Y, \magg{D})} \arrow {\tt arc(X, Y, D)}, {\tt h(X) = i}. \\
& r_{6.2}: {\tt path(X, Y, \magg{D})} \arrow {\tt path(X, Z, Dxz)}, {\tt path^{(1)}(Z, Y, Dzy)},  {\tt D=Dxz+Dzy}, {\tt h(X) = i}. \\
& r_{6.3}: {\tt path^{(1)}(X, Y, \magg{D})} \arrow {\tt path(X, Y, D)}, {\tt h(X) = i}. \\
& r_{6.4}: {\tt shortestpath(X, Y, D)} \arrow {\tt path(X, Y, D)}, {\tt h(X) = i}. \\
\end{aligned}
\end{displaymath}

\vspace{5pt}

In a bulk synchronous distributed computing model, the communication between the workers 
in each iteration can be considerably more expensive than the local computation performed by each worker due to the bottleneck of network bandwidth. 
We now investigate if we can relax this synchronization constraint at every iteration. 

Under a \emph{stale synchronous parallel} (SSP) model, a worker $i$ can use an obsolete or stale version of {\tt path}$^{(1)}$ that omits some recent updates, produced by other workers, for its local computation. 
In particular,
a worker using {\tt path}$^{(1)}$
at iteration
$c$ will be able to use all the atoms and updates generated 
from iteration $0$ to $c - s - 1$, 
$ s \geq 0 $
is a user-specified threshold for controlling the staleness.
In addition, the worker's stale {\tt path}$^{(1)}$ may have atoms or updates from 
iteration beyond 
$c - s - 1$\revision{,} i.e.\revision{,} from iteration $c - s$ to $c - 1$ (although this is not guaranteed).
The intuition behind this is that in a SSP model, a worker for its local computation should be able to see and use its own updates at every iteration, in addition to seeing and using as many updates as possible from other workers, with the constraint that any updates older than a given age are not missed. This is the \emph{bounded staleness} constraint \citeS{sspdef}.
This leads to two advantages: 

\begin{enumerate}
    \item Workers spend more time performing actual computation, rather than idle waiting for other workers to finish. This can be very helpful, when there are straggling workers present, which lag behind others in an iteration.
    In fact in distributed computing, stragglers present an acute problem since they can occur for several reasons like hardware differences \citeS{disk1}, system failures \citeS{hardfail}, skewed data distribution or even from software management issues and program interruptions caused from garbage collections or operating system noise, etc. \citeS{oseffect}.
    \item Secondly, workers can end up communicating less than under a BSP model. This is primarily because under \prem, each worker can condense several updates computed from different local iterations into a single update before eventually sending it to other workers.  
\end{enumerate}

\begin{figure*}[t]
\centering
\includegraphics[width=0.4\textwidth]{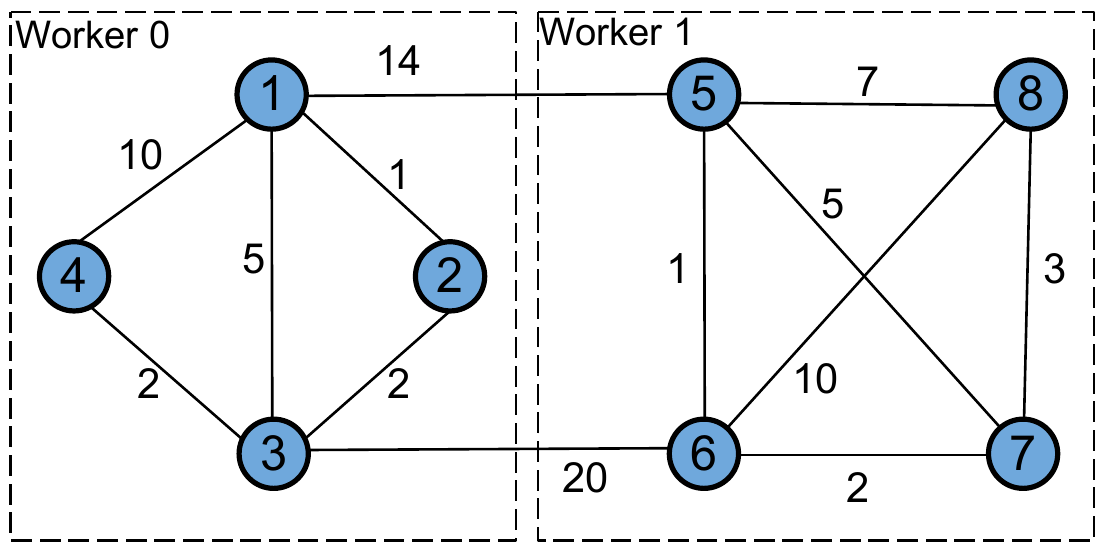}
\caption{A toy graph distributed across two workers.}
\label{fig:graph}
\end{figure*}

\begin{figure}[ht]
\centering
\includegraphics[width=0.75\textwidth]{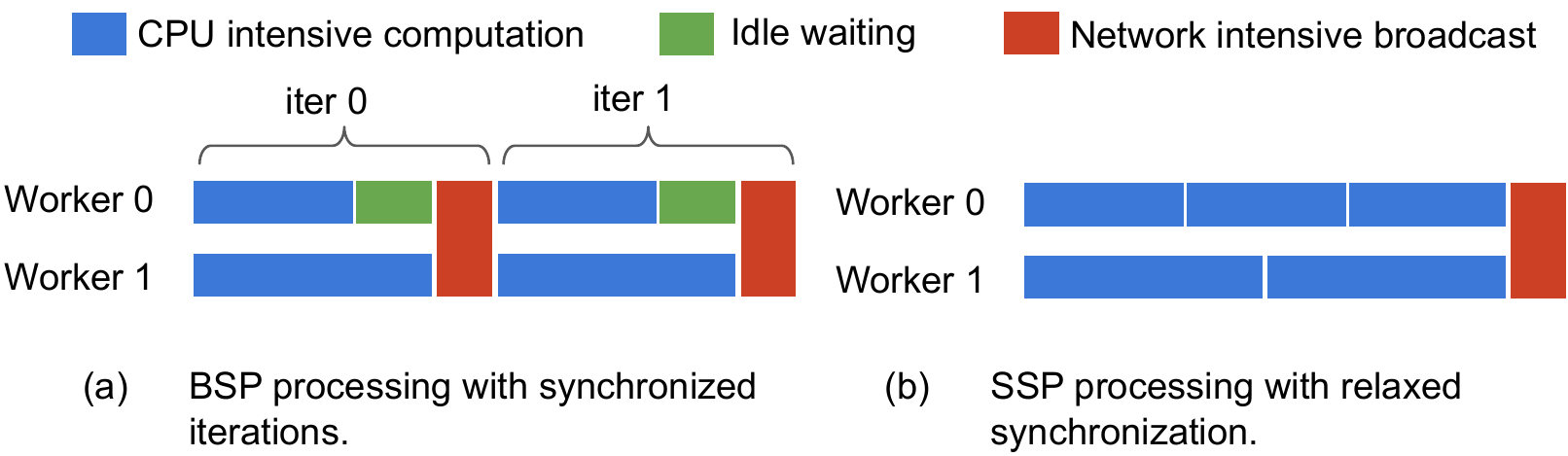}
\caption{BSP vs. SSP model for evaluating \emph{all pairs shortest path} query on two workers.}
\label{fig:mem-models}
\end{figure}

We illustrate the above advantages through an example. Figure \ref{fig:graph} shows a toy graph which is distributed across two workers: (i) all edges incident on nodes 1-4 are available on \emph{worker 0}, (ii) and the rest of the edges reside on \emph{worker 1}. \revision{Now consider the shortest path between nodes 4 and 8, given by the path 4-3-2-1-5-6-7-8, which spans across 7 hops.} The parallel program defined by rules \revision{$r_{6.1}-r_{6.4}$} with BSP processing would require at least three synchronized iterations to reach to the least fixpoint by semi-naive evaluation. Now consider \emph{worker 1} to be a straggling node that lags behind \emph{worker 0} during the computation because of hardware differences. Thus\revision{,} \emph{worker 0} spends significant time idle waiting for \emph{worker 1} to complete, as shown in Figure \ref{fig:mem-models}a. 
But in this example, the shortest path between nodes 4 and 8  changes because of two aspects: (1) the shortest path between nodes 4 and 1 changes 
and (2) the shortest path between nodes 5 and 8 changes. Both of these computations can be done independently on the two workers and 
\emph{worker 0} needs to know the eventual shortest path 
between nodes 5 and 8 calculated by \emph{worker 1} and vice versa. It is important to note that this will only work if each worker can use the most recent \emph{local} updates (newest atoms) generated by itself. 
In other words, \emph{worker 0} should be able to see the changes of the shortest path between node 4 and node 1 in every iteration (which is generated locally) and use a stale (obsolete) knowledge about the shortest path between nodes 5 and 8 (as sent by \emph{worker 1} earlier). This stale synchronization model is summarized in 
Figure \ref{fig:mem-models}b. 

In this same example, note how the minimum cost for the path between node 1 and node 4 (computed by \emph{worker 0}) changes in every iteration: (i) in the first iteration, the minimum cost was 10 given by the edge between node 1 and node 4, (ii) in the next iteration, the minimum cost drops to 7 given by the path 1-3-4 and (iii) in the third iteration the final minimum cost of 5 is given by the sequence 1-2-3-4. In a BSP model, each of this update generated in every iteration needs to be communicated to all the remaining workers. However, in a SSP model due to the advantage of this staleness, these multiple updates from different local iterations can be condensed into one most recent update, which is then sent to other workers. In other words, SSP with \prem may skip sending some updates to remote workers, thus saving communication time.    


Figure \ref{alg:ssp} formally presents the SSP processing based bottom-up evaluation plan for the non-linear all pairs shortest path example given by rules \revision{$r_{6.1}-r_{6.4}$}. If the evaluation is executed over a distributed system of $\mathcal{W}$ workers, \revision{Figure} \ref{alg:ssp} depicts the execution plan for a worker $i$. A coordinator marks the completion of the overall evaluation process by individually tracking the termination of each of the worker's task. For simplicity and clarity, we have used the naive fixpoint computation to describe the evaluation plan instead of using the optimized differential fixpoint algorithm. The $\gamma$ used in the Figure \ref{alg:ssp} denotes the $min$ constraint. Step (3) in this evaluation plan shows how worker $i$ uses stale knowledge from other workers $j$ (denoted by ${\tt path_{j}^{(r')}}$) during the recursive rule evaluation, shown by step (4). It is also important to note that in step (4), each worker $i$ is also using the most recent atoms generated by itself (denoted by ${\tt path_{i}^{(r)}}$) for the evaluation. The condition in step (6) allows each local computation on worker $i$ to reach local fixpoint or move further by at least $\mathcal{T}$ iterations. Thus, each worker $i$ can condense multiple updates generated within these $\mathcal{T}$ iterations due to \prem into a single update. Finally, step (9) ensures that if any worker falls beyond the user-defined \emph{staleness bound}, then other workers wait for it to catch up within the desired staleness level before starting their local computations again. 
We next present some theoretical and empirical results about the SSP model based bottom-up evaluation.

\begin{figure}[htbp]
\begin{mdframed}
\begin{algorithmic}[1]
\STATE ${\tt path_{i}^{(0)}(X,Y,D)} := \{({\tt X},{\tt Y},{\tt D}) | {\tt arc}({\tt X,Y,D} )\}$, ${\tt path_{j}^{(0)}(X,Y,D)} := \emptyset$ ${ \forall i \ne j}$, ${\tt r} = 0$, 
${\tt s} = 0$
\REPEAT
\STATE ${\tt path_{j}^{(r')}(X,Y,D)} := $Last received ${\tt path_{{j}}}$ by worker ${\tt i}, \forall i \ne j.$
\STATE ${\tt path_{i}^{(r+1)}(X,Y,D)} := \gamma\bigg(\hspace{2pt} \Big(\bigcup\limits_{i \ne j}{\tt path_{i}^{(r)}(X,Z,Dxz)} \bowtie {\tt path_{j}^{(r')}(Z,Y,Dzy)}\Big) \bigcup$ 

\hspace{90pt} $ \Big({\tt path_{i}^{(r)}(X,Z,Dxz)} \bowtie {\tt path_{i}^{(r)}(Z,Y,Dzy)}\Big)\biggr)$
\STATE ${\tt r} := {\tt r + 1}$, ${\tt s} := {\tt s + 1}$ 
\UNTIL {$s < \mathcal{T}$ and ${\tt path_{i}^{(r)}} \ne {\tt path_{i}^{(r-1)}}$}
\STATE ${\tt s} := 0$
\STATE Send ${\tt path_{i}^{(r)}}$ to other workers.
\IF {for any worker $j, r - r' > $ staleness bound} 
        \STATE Wait for a new update from worker $j$ before continuing  
\ENDIF
\IF {${\tt path_{i}^{(r)}} \ne {\tt path_{i}^{(r-1)}}$ or a new update has been received from worker $j$}
        \STATE repeat from Step (2)
\ELSE
        \STATE Send a ${\tt finish}$ message to coordinator.
        \IF {any new update is received from worker $j$}
                \STATE Send a ${\tt resume}$ message to coordinator.
                \STATE Repeat from step (2).
        \ENDIF
\ENDIF 
\end{algorithmic}
\end{mdframed}
\caption{SSP based bottom-up evaluation plan executed by worker $i$ for computing all pairs shortest path.}
\label{alg:ssp}
\end{figure}

%% file: ssp.tex
\section{Bottom-up Evaluation with SSP Processing}\label{ssp}
Under the SSP model, a recursive query evaluation with \prem constraints has the following theoretical guarantees: 

\begin{theorem}\label{theorem-ssp}
Let $P$ be a recursive Datalog program with ICO $T$ and let the constraint $\gamma$ be \prem to $T$ and $P$. Let $P$ have a parallel \revision{decomposable} evaluation plan that can  
be executed over $\mathcal{W}$ workers, where $Q_i$ is the program executed at worker $i$ and $T_i$ is the corresponding ICO defined over $Q_i$. 
If $\gamma$ is also \prem to $T_i$ and $Q_i$ for $1 \leq i \leq \mathcal{W}$, then:
\end{theorem}

\begin{enumerate}[(i).] 
    \item \emph{The SSP processing yields the same minimal fixpoint of $\gamma(T^{\uparrow \omega}(\emptyset))$, as would have been obtained with BSP processing}.
    \item \emph{If any worker $i$ under BSP processing requires $r$ rounds of synchronization, then under SSP processing $i$ would require $\leq r$ rounds to reach the minimal fixpoint, where $r$ rounds of synchronization in SSP model means every worker has sent at least $r$ updates.}
\end{enumerate}

\begin{proof*}
The proof is provided in \ref{app-proof}.
\end{proof*}




\subsection{SSP Evaluation of Queries without \prem Constraint}
We now consider the parallel \revision{decomposable} plan of a transitive closure query, which does not contain any aggregates in recursion. We use the same non-linear recursive example from \citeS{bigdatalog-mc}, given by rules \revision{$r_{7.1}-r_{7.3}$}, which shows the program executed by worker $i$. 
Note, in this example every worker $i$ eventually has to compute and send to other workers all {\tt tc} atoms of the form {\tt (X, Y)}, where {\tt h(X) $= i$}. Without \prem, a worker $i$ does not update its existing {\tt tc} atoms. In fact, during semi-naive evaluation of this query, at any time, only new unique atoms are appended to {\tt tc}. Thus, a SSP evaluation for the transitive closure query (without \prem) does not save any communication cost as compared to a BSP model. However, as shown in our experimental results next, the SSP model can still mitigate the influence of stragglers.

\begin{displaymath}
\begin{aligned}
& r_{7.1}: {\tt tc(X, Y)} \arrow {\tt arc(X, Y)}, {\tt h(X) = i}. \\
& r_{7.2}: {\tt tc(X, Y)} \arrow {\tt tc(X, Z)}, {\tt tc^{(1)}(Z, Y)}, {\tt h(X) = i}. \\
& r_{7.3}: {\tt tc^{(1)}(X, Y)} \arrow {\tt tc(X, Y)}, {\tt h(X) = i}. \\
\end{aligned}
\end{displaymath}

\subsection{Experimental Results}\label{exp-results}

\noindent \textbf{Setup.} We conduct our experiments on a 12 node cluster, where each node, running on Ubuntu 14.04 LTS, has an Intel i7-4770 CPU \revision{(3.40GHz, 4 cores)} with 32GB memory and a 1 TB 7200 RPM hard drive. The compute nodes are connected with 1Gbit network. Following the standard practices established in \citeS{bigdatalog}, \citeS{bigdatalog-mc}, we execute the distributed bottom-up semi-naive evaluation using an AND/OR tree based implementation in Java on each node. 
\revision{Each node executes one application thread per core.} 
We evaluate both the non-linear all pairs shortest path and transitive closure queries on a subset of the real world \emph{orkut} social network data\footnote{\url{http://snap.stanford.edu/data/com-Orkut.html}}. 
 
\vspace{9pt}

\noindent \textbf{Inducing Stragglers.} In order to study the influence of straggling nodes in a declarative recursive computation, we induce stragglers in our implementation following the strategy described in \citeS{SSPCui}. In particular, each of the nodes in our setup can be disrupted independently by a CPU-intensive background process that kicks in following a Poisson distribution and consumes at least half of the CPU resources.

\vspace{9pt}

\noindent \textbf{Analysis.} \revision{In this section, we empirically analyze the merits and demerits of a SSP model over a BSP model, by examining the following questions: (1) How does a SSP model compare to a BSP model when queries contain \prem constraints and aggregates in recursion? (2) How do these two processing paradigms compare when \prem cannot be applied? (3) And, how do the overall performances in the above scenarios change in presence and absence of stragglers? 
Table \ref{table:wprem} captures the first case with the all pairs shortest path query (where \prem is applicable), while Table \ref{table:woprem} presents the second case with the transitive closure query, which do not contain any aggregates or \prem constraints in recursion. For each of these two cases, as shown in the tables, we experimented with two different staleness values for a SSP model, both under the presence and absence of induced stragglers.   
Notably, a SSP model with bounded staleness (alternatively also called `slack' and indicated by $s$ in the tables) set as zero reduces to a BSP model. 
Tables \ref{table:wprem} and \ref{table:woprem} capture the average execution time for the query at hand under different configurations over five runs. This \emph{run time} can be divided into two components--- (1) \emph{average computation time}, which is the average time spent by the workers performing semi-naive evaluation for the recursive computation, and (2) \emph{average waiting time}, which is the average time spent by the workers waiting to receive a new update to resume computation. Tables \ref{table:wprem} and \ref{table:woprem} show the run time break down for the two aforementioned cases (with and without \prem respectively). 
}

\revision{From Tables \ref{table:wprem} and \ref{table:woprem}, it is evident that BSP processing requires the least compute time irrespective of straggling nodes. This is also intuitively true because the total recursive computation involved in a BSP based distributed semi-naive evaluation is similar to that of a single executor based sequential execution and as such a BSP model should require the least computational effort to reach the minimal fixpoint. 
On the other hand, a SSP model may perform many local computations optimistically with obsolete data using relaxed synchronization barriers, which can become redundant later on. As shown in the tables, average compute time indeed increases with higher slack indicating that a substantial amount of the work becomes unnecessary. However, as seen from both the tables, SSP plays a major rule in reducing the average wait time. This is trivially true, since in SSP processing, any worker can move ahead with local computations using stale knowledge, instead of waiting for global synchronization as required in BSP. However, note the reduction in average wait time under SSP model in Table \ref{table:wprem} (with \prem) is more significant than in Table \ref{table:woprem} (without \prem). This can be attributed to the fact that \prem with semi-naive evaluation (Section \ref{prem-overview}) 
under SSP model can batch multiple updates together before sending them, thereby saving communication cost. However, for the transitive closure query (without \prem), the overall updates sent in BSP and SSP models are similar (since no aggregates are used, semi-naive evaluation only produces new atoms, never updates existing ones). Thus, in the latter case (Table \ref{table:woprem}), the wait times between BSP and SSP models are comparable when there are no induced stragglers, whereas the wait time in SSP is marginally better than BSP when stragglers are present.
Notably, inducing stragglers obviously increases the average wait time all throughout as compared to a no straggler situation. The compute time also increases marginally in presence of stragglers, primarily because the straggling nodes take longer time to finish its computations. 
}  

\revision{Thus, to summarize based on the run times in the two tables, we see that} in absence of stragglers, the SSP model can reduce the run time of the shortest path query (with \prem constraint) by nearly 30\%. However, the same is not true for the transitive closure query, which do not have any \prem constraint. \revision{Hence, a BSP model would suffice if there are no stragglers and the query does not contain any \prem constraint.} However, in presence of stragglers \revision{or \prem constraints}, SSP model turns out to be a better alternative than BSP model, as it can lead to a execution time reduction of as high as 40\% for the shortest path query and nearly 7\% for the transitive closure query. Finally, it is also worth noting from the results that too much of a slack can also increase the query latency. Thus, a moderate amount of slack should be used in practice.

\begin{table}[h!]
\centering
\begin{tabular}{||c||c c c||c c c||} 
\hline
 \multirow{2}{*}{\hspace{2pt} \revision{Time  consumption} \hspace{2pt}} & \multicolumn{3}{c||}{\revision{No stragglers (time in sec)}} & 
 \multicolumn{3}{c||}{\revision{With stragglers (time in sec)}} \\ [0.7ex]
 & \revision{BSP \scriptsize{($s$=0)}} & \revision{SSP \scriptsize{($s$=3)}} & 
 \revision{SSP \scriptsize{($s$=6)}} & \revision{BSP \scriptsize{($s$=0)}} & 
 \revision{SSP \scriptsize{($s$=3)}} & \revision{SSP \scriptsize{($s$=6)}}  \\ [0.5ex] 
 \hline
  \revision{\emph{Avg. compute time}} &  \revision{{\bf 2224}} & \revision{ 2443} & \revision{3038} & \revision{{\bf 2664}} & \revision{2749} & \revision{3435} \\  [0.5ex]
  \revision{\emph{Avg. wait time}} &  \revision{1679} & \revision{{\bf 302}} & \revision{{\bf 408}} & \revision{2786} & \revision{{\bf 485}} & \revision{{\bf 704}} \\  [0.5ex]
 \revision{\emph{Run time}} &  \revision{3903} & \revision{{\bf 2745}} & \revision{{\bf 3446}} & \revision{5450} & \revision{{\bf 3234}} & \revision{{\bf 4139}} \\  [0.5ex]
 \hline
\end{tabular}
\caption{\revision{Comparing BSP vs. SSP model for all pairs shortest path query containing aggregates in recursion (with \prem).}}
\label{table:wprem}
\end{table}

\begin{table}[h!]
\centering
\begin{tabular}{||c||c c c||c c c||} 
\hline
\multirow{2}{*}{\hspace{2pt} \revision{Time  consumption} \hspace{2pt}} & \multicolumn{3}{c||}{\revision{No stragglers (time in sec)}} & 
 \multicolumn{3}{c||}{\revision{With stragglers (time in sec)}} \\ [0.7ex]
 & \revision{BSP \scriptsize{($s$=0)}} & \revision{SSP \scriptsize{($s$=3)}} & 
 \revision{SSP \scriptsize{($s$=6)}} & \revision{BSP \scriptsize{($s$=0)}} & 
 \revision{SSP \scriptsize{($s$=3)}} & \revision{SSP \scriptsize{($s$=6)}}  \\ [0.5ex] 
 \hline
  \revision{\emph{Avg. compute time}} & \revision{{\bf 682}} & \revision{762} & \revision{879} & \revision{{\bf 754}} & \revision{827} & \revision{921} \\ [0.5ex]
  \revision{\emph{Avg. wait time}} & \revision{367} & \revision{{\bf 345}} & \revision{{\bf 334}} & \revision{618} & \revision{{\bf 456}} & \revision{{\bf 412}} \\ [0.5ex]
 \revision{\emph{Run time}} & \revision{{\bf 1049}} & \revision{1107} & \revision{1213} & \revision{1372} & \revision{{\bf 1283}} & \revision{1431} \\ [0.5ex] 
 \hline
\end{tabular}
\caption{\revision{Comparing BSP vs. SSP model for transitive closure query containing \emph{no} aggregates in recursion (without \prem).}}
\label{table:woprem}
\end{table}

%% file: conclusion.tex
\section{Conclusion}\label{conclusion}
%
\prem facilitates and extends 
the use of aggregates in recursion, and this enables
a wide spectrum of graph and data mining algorithms to be expressed efficiently in declarative languages. In this paper,  we explored various
improvements to scalability via paralled execution with \prem. In fact, \prem can be easily integrated with most of the current generation Datalog engines like \emph{BigDatalog}, \emph{Myria}, \emph{BigDatalog-MC}, \emph{SociaLite}, \emph{LogicBlox},
irrespective of their architecture differences and varying synchronization constraints. 
Moreover, in this paper, we have shown  that \prem brings additional benefits to
the parallel evaluation of recursive queries. For that, we 
established the necessary theoretical framework that allows bottom-up recursive computations to be carried out over stale synchronous parallel model---in addition to the synchronous or completely asynchronous computing models studied in the past. These theoretical developments lead us to the conclusion, confirmed by initial experiments, that the parallel execution of non-linear queries with \prem constraints can be  expedited with a stale synchronous parallel (SSP) model. This model is also useful in the absence of
 \prem constraints, where bounded staleness may not reduce  communications, but it nevertheless mitigates the impact of stragglers. Initial experiments performed on a real-world dataset  confirm the theoretical results, and are quite promising, paving the way toward future research 
 \revision{in many interesting areas, where declarative recursive computation under SSP processing can be quite advantageous.
 For example, declarative advanced stream reasoning systems \citeS{astro-cikm}, supporting aggregates in recursion, can adopt distributed SSP model to query evolving graph data, especially when one portion of the network changes more rapidly as compared to others. SSP models under such scenario offer the flexibility to batch multiple network updates together, thereby reducing the communication costs effectively.}
 
 \revision{Finally, it is important to note that the methodologies developed here can also be applied to other declarative logic based systems beyond Datalog, like in SQL-based query engines \citeS{rasql}, which also use semi-naive evaluation for recursive computation. 
 In addition, the SSP processing paradigm can also be adopted in many  state-of-the-art graph-centric platforms such as Pregel \citeS{pregel} and GraphLab \citeS{graphlab}. 
 These modern graph engines use a vertex-centric computing model \citeS{vertexProgramming}, which enforces a strong consistency requirement among its model variables under the ``Gather-Apply-Scatter'' abstraction. Consequently, this makes the synchronization cost for these graph frameworks similar to that of standard BSP systems.
 Thus, for many distributed graph computation problems involving aggregators (like shortest path queries), SSP model, as demonstrated in this paper, can be quite useful for these graph based platforms. 
 }
%
%

%% file: app.tex
\newpage
\appendix
\section{SSP processing based recursive computation with \prem}\label{app-proof}

\begin{definition}{($\gamma$-Cover).}
Let $P$ be a positive recursive Datalog program with $T$ as its corresponding ICO. Let a constraint $\gamma$ be defined over the recursive predicate on a set of $k$ group-by arguments, denoted by $G_1, G_2, ..., G_k$ with the cost-argument denoted as $C$. Let $\gamma$ be also \prem to $T$ and $P$. Let there be two sets $S_1$ and $S_2$, both of which contain tuples of the form $\{(g_1, g_2, ..., g_k, c)|g_i \in G_i \forall 1 \leq i \leq k, c \in \mathbb{R} \}$, where $\mathbb{R}$ represents the set of real numbers. Now, $S_1$ is defined as the  $\gamma$-cover for $S_2$, if for every tuple $t \in S_2$, there exists only one tuple $t' \in S_1$ such that (i) $t'[G] = t[G]$ and (ii) $\gamma(t'[C], t[C]) = t'[C]$.
\end{definition}

It is important to note from the above definition that if $S_1$ is the $\gamma$-cover for $S_2$, then there can exist a tuple $t \in S_1$, such that $t[G] \ne t'[G]$ $\forall t' \in S_2$ but the converse is never true.

\vspace{9pt}

\noindent \textbf{Lemma 2.} 
Let $P$ be a recursive Datalog program, $T$ be its corresponding ICO and let the constraint $\gamma$ be \prem to $T$ and $P$, resulting in the constrained ICO $T_{\gamma}$. Now, for any pair of positive integers $m, n$, where $m \geq n$, $T_\gamma^{\uparrow m}(\emptyset)$ is a $\gamma$-cover for $T_\gamma^{\uparrow n}(\emptyset)$.

\begin{proof*}
This directly follows from the fact that any atom in $T_\gamma^{\uparrow n}(\emptyset)$ with cost $c$ can only exist in $T_\gamma^{\uparrow m}(\emptyset)$ with updated cost $c'$, if $c=c'$ or $\gamma(c, c') = c'$. Note if $c=c'$, then $\gamma(c, c') = c'$ is trivially true.
\end{proof*}

\vspace{9pt}

\noindent \textbf{Lemma 3.} 
Let $P$ be a recursive Datalog program with ICO $T$ and let the constraint $\gamma$ be \prem to $T$ and $P$. Let $P$ also have a parallel \revision{decomposable} evaluation plan that can  
be executed over $\mathcal{W}$ workers, where $Q_i$ is the program executed at worker $i$ and $T_i$ is the corresponding ICO defined over $Q_i$. 
Let $\gamma$ be also \prem to $T_i$ and $Q_i$, for $1 \leq i \leq \mathcal{W}$. After $r$ rounds of synchronization ($r$ rounds of synchronization in SSP model means every worker has sent at least $r$ updates), if $I_b$ and $I_s$ denote the interpretation of the recursive predicate under BSP and SSP models respectively for any worker $i$, then $I_s$ is a $\gamma$-cover for $I_b$. 

\begin{proof*}
In a SSP based fixpoint computation, any worker $i$ can produce an atom in three ways: 
\end{proof*}

\begin{enumerate}[(1)]
    \item From local computation not involving any of the updates sent by other workers.
    \item From a join with a new atom or an update sent by another worker $j$.
    \item From both cases (1) and (2) together. 
\end{enumerate}

Now, consider the base case, where before the first round of synchronization (i.e.\revision{,} at the $0^{th}$ round) each worker performs only local computation, since it has not received/sent any update from/to any other worker. Since, in a SSP model, each local computation may involve multiple iterations (as shown in step (6) in Figure \ref{alg:ssp}), $I_s$ is trivially a $\gamma$-cover for $I_b$ (from lemma 2).

We next assume this hypothesis (lemma 3) to be true for some $r \geq 0$. Under this assumption, we find that each worker $i$ in SSP model for its fixpoint computation operates based on the information from its own $I_s$ and from the ones sent by other workers after the $r^{th}$ round of synchronization. And since each of this $I_s$ involved is a $\gamma$-cover for the corresponding $I_b$ (when compared against the BSP model), the aforementioned cases (1)-(3) will also produce a $\gamma$-cover for the $(r+1)^{th}$ synchronization round. 

Hence, by principle of mathematical induction, the lemma holds for all $r \geq 0$.  

\vspace{9pt}

\begin{customthm}{2}
Let $P$ be a recursive Datalog program with ICO $T$ and let the constraint $\gamma$ be \prem to $T$ and $P$. Let $P$ have a parallel \revision{decomposable} evaluation plan that can  
be executed over $\mathcal{W}$ workers, where $Q_i$ is the program executed at worker $i$ and $T_i$ is the corresponding ICO defined over $Q_i$. 
If $\gamma$ is also \prem to $T_i$ and $Q_i$, for $1 \leq i \leq \mathcal{W}$, then:
\end{customthm}

\begin{enumerate}[(i).] 
    \item \emph{The SSP processing yields the same minimal fixpoint of $\gamma(T^{\uparrow \omega}(\emptyset))$, as would have been obtained with BSP processing}.
    \item \emph{If any worker $i$ under BSP processing requires $r$ rounds of synchronization, then under SSP processing $i$ would require $\leq r$ rounds to reach the minimal fixpoint, where  $r$ rounds of synchronization in SSP model means every worker has sent at least $r$ updates.}
\end{enumerate}

\begin{proof*}
Theorem \ref{theorem-prem} guarantees that the BSP evaluation of the datalog program with \prem will yield the minimal fixpoint of $\gamma(T^{\uparrow \omega}(\emptyset))$. 
Note that in the SSP evaluation, for every tuple $t$ produced by a worker $i$ from the program $Q_i$, $t \in T^{\uparrow \omega}(\emptyset)$. In other words, 
if $I$ represents the final interpretation of the recursive predicate under SSP evaluation, 
then $I \subseteq T^{\uparrow \omega}(\emptyset)$ i.e. $I$ is bounded. It also follows from lemma 3, that $I$ is a $\gamma$-cover for the final interpretation of the recursive predicate under BSP evaluation i.e. $I$ is a $\gamma$-cover for $\gamma(T^{\uparrow \omega}(\emptyset))$. Since, $\gamma(T^{\uparrow \omega}(\emptyset))$ is the least fixpoint under the $\gamma$ constraint, we also get $\gamma(T^{\uparrow \omega}(\emptyset)) \subseteq I$, as atoms in $\gamma(T^{\uparrow \omega}(\emptyset))$ must have identical cost in $I$. 

Thus, we can write the following equation based on the above discussion, 
\begin{equation}\label{eqa1}
\gamma(T^{\uparrow \omega}(\emptyset)) \subseteq I \subseteq T^{\uparrow \omega}(\emptyset) 
\end{equation}

Also recall, since $\gamma$ is \prem to each $T_i$ and $Q_i$, under the SSP evaluation, each worker $i$ also applies $\gamma$ in every iteration in its fixpoint computation (step (4) in Figure \ref{alg:ssp}). Thus, we have,
\begin{equation}\label{eqa2}
  I \subseteq  \gamma(T^{\uparrow \omega}(\emptyset))
\end{equation}

Combining equations (\ref{eqa1}) and (\ref{eqa2}), we get $I = \gamma(T^{\uparrow \omega}(\emptyset)$. Thus, the SSP evaluation also yields the same minimal fixpoint as the BSP model. 

\vspace{3pt}

Since, the interpretation of the recursive predicate in the least model obtained from BSP evaluation is identical to that in the least model obtained from SSP processing, it directly follows from lemma 3, that the number of synchronization rounds required by worker $i$ in SSP evaluation will be at most $r$, where $r$ is the number of rounds $i$ takes under BSP model.  
\end{proof*}